\documentclass[11pt]{article}
\pdfoutput=1
\usepackage{jheppub}
\usepackage{amssymb,amsfonts}
\usepackage{mathrsfs}
\usepackage{amsmath}
\usepackage{cleveref}
\let\savenumberline\numberline
\def\numberline#1{\savenumberline{#1.}}
\makeatletter
\renewcommand{\@seccntformat}[1]{\csname the#1\endcsname.\,\,}
\makeatother

\newcommand{\Z}{{\bf Z}}

\newcommand{\CB}{{\cal B}}

\newcommand{\CH}{{\cal H}}

\newcommand{\CL}{{\cal L}}

\newcommand{\SB}{{\mathscr B}}

\newcommand{\MR}{{\mathbb R}}

\renewcommand{\tilde}[1]{\widetilde{#1}}
\renewcommand{\hat}[1]{\widehat{#1}}
\newcommand{\be}{\begin{equation}}
\newcommand{\ee}{\end{equation}}
\newcommand{\bea}{\begin{eqnarray}}
\newcommand{\eea}{\end{eqnarray}}

\newcommand\secref[1]{{\S\ref{#1}}}
\newcommand\appref[1]{{Appendix~\ref{#1}}}

\makeatletter
\def\@fpheader{\relax}
\makeatother
\usepackage{graphicx}
\usepackage{latexsym}

\title{\ \vspace{1.5in} \\ \hbox{Tropical Branes}}
\author{Emil Albrychiewicz, Andr\'{e}s Franco Valiente, and Vi Hong}
\affiliation{\medskip
Berkeley Center for Theoretical Physics and Department of Physics\\
University of California, Berkeley, CA, 94720-7300, USA\medskip\\
Theoretical Physics Group, Lawrence Berkeley National Laboratory\\
Berkeley, CA 94720-8162, USA}
\emailAdd{ealbrych@berkeley.edu}
\emailAdd{andresfranco@berkeley.edu}
\emailAdd{vihong14@berkeley.edu}
\abstract{We investigate canonically quantized open string solutions associated to the analytically continued action for the recently proposed  tropical limit of topological A-type models, \textit{tropological sigma models}, with various tropical versions of boundary conditions. These solutions naturally give rise to a non-relativistic counterpart of branes, which we name \textit{tropical branes}. We provide a worldsheet description of these tropical branes. We find that instead of the usual string spectrum of infinitely many coupled harmonic oscillators, the Hamiltonian for tropical branes describes an infinite tower of increasingly massive asymptotic string states. This results supports the claim that a tropicalization of the worldsheet formulation of string theory gives a viable description for the asymptotic behavior of string amplitudes and their non-equilibrium extensions.
}
\begin{document}
\maketitle

\section{Introduction}
\label{sec:Intro}

In \cite{trsm}, a new class of cohomological quantum field theories was constructed by taking an ultralocal limit, known as the \textit{tropicalization} \cite{msintro, rau, mikhalkinrau, litvinov}, of the topological A-type sigma models \cite{ewtsm}. These theories were coined \textit{tropological sigma models}. Unlike the relativistic A-model, whose worldsheet $\Sigma$ is equipped with a complex structure, these tropological models are described by degenerate, nilpotent endomorphisms of the worldsheet tangent bundle $T\Sigma$, or equivalently by worldsheet foliations. The symmetries that preserve this geometric structure are nonrelativistic in nature and manifest as foliation-preserving diffeomorphisms, similar to those found in other theories that admit a worldsheet formulation such as Carrollian, non-vibrating, and ambitwistor string theories \cite{ziqi, Blair:2023noj}. 

The physical motivation to study the tropical limit of topological sigma models resides in the attempt to construct a non-equilibrium string worldsheet perturbation theory based on the Schwinger-Keldysh formalism. This formalism involves a doubling of the fields such that we have degrees of freedom propagating forward in time on one branch and fields that propagate backwards in time on a backwards branch. This results in a double decomposition of perturbative amplitudes. In \cite{neq, ssk, keq}, it was argued that unlike conventional relativistic quantum field theories, the string-theoretic reformulation of the Schwinger-Keldysh contour results in a triple decomposition of the worldsheet of the form
\[
\Sigma = \Sigma^{+} \cup \Sigma^{\wedge} \cup \Sigma^{-},
\]
with \(\Sigma^{+}\) and \(\Sigma^{-}\) corresponding to the forward and backward branches, respectively, and the ``wedge region” \(\Sigma^{\wedge}\) connecting them at the turnaround. Remarkably, this wedge region is a bona fide two-dimensional region with its own genus expansion and arbitrarily complex topology and was argued to be geometrically highly anisotropic. In this setting, the worldsheet develops non-relativistic foliations, wherein leaves connect boundary points on \(\Sigma^{+}\) with corresponding boundary points on \(\Sigma^{-}\). Naturally, one is led to ask: how can this wedge region be constructed explicitly for worldsheet path integral computations?

In \cite{trsm}, it was shown that one is able to construct non-relativistic worldsheet theories that exhibit a notion of anisotropy similar to the wedge region by utilizing methods of tropical geometry. These tropicalized worldsheets not only display the predicted non-relativistic foliation structure but also suggest that the target space geometry itself becomes tropicalized. From the perspective of non-equilibrium string perturbation theory, we expect their analytic continuations to describe the dynamics of string worldsheet which have degenerated into long tubes that represent on-shell asymptotic string states and are therefore physically equivalent to an infinite tower of free particles. We find supporting evidence for this statement in this paper.

In tropical geometry, the underlying complex numbers field is deformed into what is known as a \textit{tropical semiring}, where the usual operations of addition and multiplication are replaced by the max and addition operations, respectively. The tropological sigma models can be formulated through the use of standard BRST cohomological methods in terms of a conventional path integral that localizes on degenerate geometric objects, which describe the tropical localization equations. An explicit action was constructed in \cite{trsm}, where it was shown to have a consistent analytic continuation into real time such that the anisotropic\footnote{Anisotropic, non-relativistic and tropical are used interchangeably in this context.} conformal symmetries are preserved and the theory remains unitary. We review the main points of this construction in \secref{sec:Review} and refer readers for more details to the original work \cite{trsm}. 

D-branes play a central role in the non-perturbative sector of string theory. Naturally, this raises the question of how branes manifest in a tropical string theory. The exploration of D-brane boundary conditions in both physical and topological string theories was first initiated in \cite{Horava:1989vt, Horava:1993ts, Witten:1992fb}. Specifically, we examine the analytically continued tropological sigma model on a worldsheet with boundaries. We consider the  strip
\begin{align}
\label{eqn:BdryWorldsheet}
    \Sigma =  \MR\times[0,\pi] ,
\end{align}
where the spatial coordinate $r$ is bounded in the interval $[0,\pi]$, and we regard this as an open tropicalized string theory with boundaries at $r=0$ and $r=\pi$. We discuss the form of the two types of physically distinct boundary conditions, which we still label as Dirichlet (D) and Neumann (N) by considering variations of the tropicalized action. These boundary conditions are different from the relativistic ones, modifying the mode expansions of the equations of motion. The generators of the anisotropic conformal symmetries also lead to a new type of boundary algebra, discussed in \secref{sec:Algebra}. Interestingly, these solutions are reminiscent of the Carrollian D-branes found in \cite{Bergshoeff:2020xhv, Kluson:2022jxh, Bergshoeff:2023rkk}. Finally, in \secref{sec:OpenStrings}, we discuss the open strings solutions with different boundary conditions and the canonical quantization of this theory.

\section{Review of Tropological Sigma Models}
\label{sec:Review}

The standard relativistic sigma models are quantum field theories that describe the dynamics of maps $\Phi: \Sigma \rightarrow M$, where $\Sigma$ is a Riemann surface and $M$ is a complex manifold equipped with a metric. This theory can be supersymmetrized in various ways. For our purposes, we want to work with a $\mathcal{N}=(2,2)$ SUSY sigma model where the target space $M$, is a Kähler manifold. In this context, there exist two well-known topological twists that reduce the number of supersymmetries by a half, called the A-model and the B-model \cite{ewmirror}. We primarily work with the A-model in this paper. In this paper, we replace $\Sigma$ and $M$ with their tropical counterparts and we review the idea of the tropical limit in details after the discussion of some generalities about the sigma models. 

To establish notation, we denote the worldsheet complex structure as $\hat{\varepsilon}$, worldsheet coordinates as $\sigma^\alpha$, and the worldsheet metric as $\hat{g}$. Locally, the maps $\Phi$ are given by the coordinate representatives $Y^i(\sigma)$. We use a hat to distinguish these geometric structures from their tropicalized counterpart, consistent with the notation introduced in \cite{trsm}. For this paper, we assume that $M$ is a smooth manifold of real dimension 2 equipped with an almost complex structure $\hat{J}$. 

By construction, the A-model localizes on pseudoholomorphic maps which are given by
\begin{equation}
\label{eqn:LocEqnOld}
\bar{\partial}_{\hat{\varepsilon}, \hat{J}} \Phi=d \Phi+\hat{J} \circ d \Phi \circ \hat{\varepsilon}=0 .
\end{equation}
Prior to the tropical limit, one can use the defining property of a complex structure $\hat{J}^2=-1$, to write this equivalently as $\hat{J} \circ d \Phi=d \Phi \circ \hat{\varepsilon}$. Locally, these equations are then written as
\begin{align}
\label{eqn:LocEqn}
    \hat{E}_\alpha^{\;\;i}=\hat{\varepsilon}_{\alpha}^{\;\;\beta}\partial_\beta Y^i - \hat{J}_j^{\;\;i}\partial_\alpha Y^j=0.
\end{align}
The tropical limit of \eqref{eqn:LocEqnOld} is not equivalent to the tropical limit of \eqref{eqn:LocEqn}, due to the fact that the defining property of the complex structures is lost under tropicalization and is instead replaced by a nilpotency condition $J^2=0$, $\varepsilon^2=0$, as we will explain below. Starting off with  \eqref{eqn:LocEqnOld} leads to a trivial tropical limit, where no useful physics can be extracted from the tropical localization equations, i.e., it is crucial to start with \eqref{eqn:LocEqn}.

We now review the tropicalization procedure. We begin by noting a main difference between our approach to tropicalization and the one commonly used in the mathematical literature, where the manifold is reduced to a middle-dimensional tropical variety. In our approach, all dimensions of the original manifold are preserved under the limit. In order to implement this into the path integral, the tropical limit is best understood as a Maslov (also called Litvinov-Maslov) dequantization \cite{msintro, litvinov, viro, virohyper} of the underlying coordinates. This singular deformation of the coordinates then induces corresponding degenerate deformations of other geometric structures such as the complex structure and metric, which naturally act on the tangent bundle. Vector bundles not associated with the tangent bundle have to be supplemented with an additional prescription based on the context.

Since we have assumed the target space dimension is two real-dimensional, we have integrable complex structures on both $\Sigma $ and $M$, and hence we can construct local complex coordinates $(z,\bar{z})$ and $(Z,\bar{Z})$ on these spaces, respectively. Then we parametrize them by introducing Viro's subtropical deformation of the complex numbers \cite{viro},
\begin{align}
\label{eqn:ViroSubTrop}
    z=\exp\left\{\frac{r}{\hbar}+i\theta\right\}, \quad Z=\exp\left\{\frac{X}{\hbar}+i\Theta\right\},
\end{align}
where $(r,\theta)$ and $(X,\Theta)$ are adapted worldsheet and target space polar coordinates. This prescription appears to be dependent on the coordinate system used but as explained in \cite{trsm} there is a new symmetry emerging (see \eqref{eqn:AlphaSym1} and \eqref{eqn:AlphaSym2}), which can be used to express coordinate changes as local gauge transformations. Moreover, transition functions between different local coordinate systems on the tropicalized manifold are restricted to be piece-wise linear. It is natural to identify the real coordinates as $\sigma^\alpha=(r,\theta)$ and $Y^i=(X,\Theta)$. The tropical limit is achieved by taking $\hbar\rightarrow 0$ of objects that depend on \eqref{eqn:ViroSubTrop}. Once we take the tropical limit, the complex structures $\hat{\varepsilon}$ and $\hat{J}$ reduce to
\begin{align}
\label{eqn:TropJor}
    \varepsilon_{\alpha}^{\;\;\beta}d\sigma^\alpha \otimes \frac{\partial}{\partial \sigma^\beta}=dr \otimes\frac{\partial}{\partial\theta}, \quad J_{i}^{\;\;j}dY^i\otimes\frac{\partial}{\partial Y^j}=dX\otimes\frac{\partial}{\partial \Theta},
\end{align}
up to an overall $\hbar$ dependent renormalization. From the 2-dimensional perspective, these are now degenerate nilpotent endomorphisms
\begin{align}
    \varepsilon^2=0, \quad J^2=0.
\end{align}
The matrix representations of these nilpotent endomorphisms have a canonical Jordan block form, and for this reason, we refer to them as Jordan structures. These Jordan structures define, pointwise, a filtration of vector spaces. For vectors $v\in F^1V$ annihilated by $\varepsilon$ we have
\begin{align}
\label{eqn:foliation}
    0\subset F^1V \subset F^2V = V,
\end{align}
where $F^1V$ is one dimensional. The filtration also defines an integrable distribution, which gives rise to a unique foliation, as our spaces are two real-dimensional. In the adapted coordinates, the leaves of the foliations of both the worldsheet and the target space are parametrized by the periodic coordinates $\theta$ and $\Theta$ respectively.

In the tropical limit, the metric  tensor $\hat{g}$ becomes a degenerate bilinear form. In adapted coordinates, one finds
\begin{align}
    g_{\alpha\beta}d\sigma^{\alpha}\otimes d\sigma^{\beta}=dr\otimes dr.
\end{align}
Hence, the metric tensor no longer has an inverse. The tropical limit of the inverse metric tensor $\hat{h}$ correspondingly leads to
\begin{align}
    h^{\alpha\beta}\frac{\partial}{\partial\sigma^{\alpha}}\otimes\frac{\partial}{\partial\sigma^{\beta}}=\frac{\partial}{\partial\theta}\otimes \frac{\partial}{\partial\theta}.
\end{align}
Consequently, $g$ and $h$ now satisfy what is known as a "mutual invisibility" condition $g_{\alpha\beta}h^{\beta\gamma}=h^{\alpha\beta}g_{\beta\gamma}=0$. Naturally, one can ask what gauge symmetries are associated with preserving either the pair of degenerate metrics or, equivalently, the Jordan structure. These transformations can be interpreted as a nonrelativistic conformal symmetry since they satisfy an algebra of Weyl rescalings that preserve the degenerate metric $g$.
In fact, it turns out that the natural gauge symmetries on the tropicalized worldsheet are foliation-preserving diffeomorphisms
\begin{equation}
\begin{aligned}
& \widetilde{r}=\widetilde{r}(r), \\
& \widetilde{\theta}=\widetilde{\theta}_0(r)+\theta \partial_r \widetilde{r}(r).
\end{aligned}
\end{equation}
In the adapted coordinates, the generators associated to these foliation preserving diffeomorphisms are
\begin{align}
\label{eqn:ConfSymGen1}
    \delta r &= f(r), \\
\label{eqn:ConfSymGen2}
    \delta \theta &= F(r)+\theta\partial_r f(r).
\end{align}
Here $f(r)$ and $F(r)$ are two real, arbitrary projectable (ie. only depend on $r$) differentiable functions that vary along the foliation. 

In the tropical limit, the localization equations are obtained by replacing the complex structures with Jordan structures in \eqref{eqn:LocEqn}. In adapted coordinates, the localization equations take the form
\begin{align}
\label{eqn:LocEqnAdaptedCoord}
    E_r^{\;\;X}=\partial_\theta X=0, \quad E_r^{\;\;\Theta}=\partial_\theta\Theta-\partial_r X=0, \quad E_{\theta}^{\;\;X}=0, \quad E_{\theta}^{\;\;\Theta}=-\partial_\theta X=0. 
\end{align}
Within a local neighborhood of the worldsheet, one can write down a general solution
\begin{align}
    X(r,\theta)&=X_0(r), \\
    \Theta(r,\theta)&=\Theta_0(r)+\theta\partial_rX_0(r),
\end{align}
where $X_0(r)$ and $\Theta_0(r)$ are differentiable functions of $r$ only. 

The tropical localization equations are additionally invariant under a new type of symmetry that is not present in the relativistic case. Following the nomenclature of \cite{trsm}, we call it an $\alpha$ symmetry. The infinitesimal variations of fields $X, \Theta$ that preserve localization equations \eqref{eqn:LocEqnAdaptedCoord} are
\begin{align}
\label{eqn:AlphaSym1}
    \delta X&=\alpha_1(r),\\
\label{eqn:AlphaSym2}    
    \delta \Theta&=\alpha_0(r)+\theta\partial_r\alpha_1(r).
\end{align}
In the quantization procedure, the $\alpha$ symmetry is interpreted as a gauge symmetry that one has to fix to impose the appropriate momenta constraints. 

In order to construct a representative for the action, we construct sections that transform naturally under foliation preserving diffeomorphisms. If we postulate that fields $X$, and $\Theta$ transform as scalars under this symmetry, then the transformations that keep the localization equations \eqref{eqn:LocEqnAdaptedCoord} invariant are
\begin{align}
    \delta X&=f(r)\partial_r X+\left(F(r)+\theta\partial_r f(r)\right)\partial_\theta X,\\
    \delta \Theta&=f(r)\partial_r\Theta+\left(F(r)+\theta\partial_r f(r)\right)\partial_\theta \Theta.
\end{align}

We wish to construct a cohomological field theory associated to the moduli space of these tropical localization equations by writing an action whose bosonic component minimizes precisely on the localization equations. Following the method set by \cite{ewcoho}, we use standard BRST cohomological methods in order to construct a BRST invariant path integral.

In order to avoid naive divergences associated with taking the tropical limit of the metric used to construct Lagrangian, one can utilize tensor densities in order to construct a representative for the action directly from a gauge fixing fermion. We introduce the following tensor densities:  $\CB^{\alpha}_{\;\;i}$, a bosonic auxiliary field, together with an antighost superpartner $\chi^{\alpha}_{\;\;i}$, which are put together into an antighost BRST multiplet.
\begin{align}
    \{Q,\chi^{\alpha}_{\;\;i}\}&=\CB^{\alpha}_{\;\;i}, \\
    [Q,\CB^{\alpha}_{\;\;i}]&=0,
\end{align}
with a nilpotent graded differential, known as the  BRST charge $Q$. This BRST charge when acted on fields $Y^i$ reflects an underlying topological symmetry $[Q,Y^i]=\psi^i$, with ghosts fields $\psi^i$. Then, the BRST invariant action can be written as
\begin{align}
    \int_\Sigma d^2\sigma \CB^{\alpha}_{\;\;i}E_{\alpha}^{\;\;i},
\end{align}
up to an addition of BRST exact terms that preserve both the non-degeneracy and the BRST cohomology class of the path integral. The auxiliary field $\CB^{\alpha}_{\;\;i}$ is, by construction, gauge invariant. There is an additional gauge freedom \cite{trsm}, using which we are able to set half of the components of $\CB^{\alpha}_{\;\;i}$ to zero, as follows. There is no choice that would be covariant both on $\Sigma$ and $M$ and therefore, in the same way as in \cite{trsm}, we choose to preserve worldsheet covariance and set
\begin{align}
    \CB^{\alpha}_{\;\;X}=0.
\end{align}
We denote the leftover components as
\begin{align}
    \CB^{r}_{\;\;\Theta}=B, \quad \CB^{\theta}_{\;\;\Theta}=-\beta.
\end{align}
In this gauge, and after integrating out a term quadratic in $B$, the bosonic part of the action is
\begin{align}
\label{eqn:TropAction}
    S=\int_{\Sigma} dr d\theta\left\{\frac{1}{2}(\partial_\theta\Theta-\partial_r X)^2+\beta\partial_\theta X\right\}.
\end{align}
One can notice that we did not integrate out the Lagrange multiplier field $\beta$. The reason is that we cannot add a $\beta^2$ term that is consistent with the anisotropic conformal invariance. 

\section{Analytic Continuation of Tropological Sigma Models}
\label{sec:AnalyticCont}
Having reviewed the main aspects of the recently proposed tropological sigma models, we would like to study the theory with propagating local degrees of freedom. As in the previous work \cite{trsm}, we consider the real worldsheet time, which is now a non-periodic worldsheet coordinate $\theta$, to run along the leaves of the foliation, discussed around \eqref{eqn:foliation}. Hence, we consider an analytic continuation into real time by $\theta\rightarrow i t$. However, this requires careful consideration as a naive continuation will make the action complex-valued, violating unitarity.

One approach, which we do not pursue in this paper but mention due to its relevance for Carrollian theories, is to treat $\theta$ already as a real time but then the theory's energy is not bounded from below or above \cite{trsm}. To cure the unboundedness, one has to perform a conformal reduction that reduces the degrees of freedom. These reductions can be achieved consistently by gauging a subgroup of the aforementioned $\alpha$ symmetry \eqref{eqn:AlphaSym1}-\eqref{eqn:AlphaSym2}. We consider two distinct subgroups that lead to two different theories with actions (see \cite{trsm} for more details)
\begin{align}
    S_1&=\int_{\Sigma} dr d\theta \left\{\frac{1}{2}(\partial_\theta\Theta)^2\right\}, \\ S_2&=\int_{\Sigma} dr d\theta \left\{\frac{1}{2}(\partial_r X)^2+\beta\partial_\theta X\right\}.
\end{align}
In the literature, these are respectively referred to as the electric and magnetic sectors of Carrollian scalar field theories \cite{Duval:2014uoa} and their quantization was discussed in \cite{deBoer:2023fnj}. 

In this paper, we take another approach, which is to add to the Lagrangian of \eqref{eqn:TropAction} a topological invariant given by the pull-back of the two form $dX\wedge d\Theta$ to $\Sigma$:
\begin{align}
    \nonumber
    \tilde{\CL}&=\CL+\partial_rX\partial_\theta\Theta-\partial_r\Theta\partial_\theta X \\ 
    \label{eqn:ModifiedAction}
    &=\frac{1}{2}(\partial_\theta\Theta)^2+\frac{1}{2}(\partial_r X)^2+(\beta-\partial_r\Theta)\partial_\theta X.
\end{align}
We can arrive at this alternative Lagrangian action by directly taking the tropical limit on the bosonic part of the action of the topological A-model,
\begin{align}
    \int_{\Sigma}d^2z\left\{\frac{1}{2}g_{ij}\partial Y^i\bar{\partial}Y^j\right\},
\end{align}
instead of the localization equations \eqref{eqn:LocEqn}.  We apply Viro's subtropical deformation \eqref{eqn:ViroSubTrop} to the integral measure, metric and fields. Working in the adapted coordinate system $(r,\theta)$ for the worldsheet and $(X,\Theta)$ for the target space, we obtain
\begin{align}
    \int dr d\theta\left\{\frac{1}{2}(\partial_r X)^2+\frac{1}{2}(\partial_\theta \Theta)^2+\frac{1}{\hbar^2}(\partial_\theta X)+\frac{\hbar^2}{2}(\partial_r \Theta)^2\right\},
\end{align}
up to an overall normalization constant. In the limit $\hbar\rightarrow 0$, the third term is divergent. We remove this divergence by introducing an auxiliary field $\xi$ and apply the Hubbard-Stratonovich transformation. This step is inspired by a derivation in \cite{ziqi} of a Galilean limit of the Polyakov action. After the transformation, the action reads
\begin{align}
    \int dr d\theta\left\{\frac{1}{2}(\partial_r X)^2+\frac{1}{2}(\partial_\theta \Theta)^2+\xi\partial_\theta X+\frac{\hbar^2}{2}(\xi^2+(\partial_r \Theta)^2)\right\}.
\end{align}
We may now take the $\hbar\rightarrow 0$ limit without running into any divergences. With the identification $\xi=\beta-\partial_r\Theta$, the following action
\begin{align}
    \int dr d\theta\left\{\frac{1}{2}(\partial_r X)^2+\frac{1}{2}(\partial_\theta \Theta)^2+(\beta-\partial_r\Theta)\partial_\theta X\right\},
\end{align}
now recovers \eqref{eqn:ModifiedAction}. 

It is now clear how to analytically continue \eqref{eqn:ModifiedAction} with $\theta\rightarrow it$. We keep the fields $X$ and $\Theta$ as real valued. In addition to continuing $\theta\rightarrow it$, we analytically continue the shifted field $\beta'=\beta-\partial_r\Theta$ as $\beta'=i\SB$ wit $\SB$ being real valued. After this analytic continuation, the action becomes
\begin{align}
\label{eqn:TropStringsAction}
    S=\int_{\Sigma} dtdr\left\{\frac{1}{2}\left(\partial_t \Theta\right)^2-\frac{1}{2}\left(\partial_r X\right)^2+(\beta-\partial_r\Theta)\partial_t X\right\}.
\end{align}
This action preserves full conformal invariance. It is also unitary with the energy being bounded from below. We refer to\eqref{eqn:TropStringsAction} as the tropical strings action.

The Euler-Lagrange equations of this action are
\begin{align}
    \label{eqn:Eom1}
        0&=\partial_r^2 X+\partial_t\partial_r\Theta-\partial_t\beta, \\ 
    \label{eqn:Eom2}
        0&=\partial_t^2\Theta-\partial_r\partial_t X, \\
    \label{eqn:Eom3}
        0&=\partial_t X.
\end{align}
We use the $\alpha$ symmetry to set $\beta=0$ (the Lagrange multiplier field $\beta$ has its own $\alpha$ symmetry: $\delta\beta=\alpha_\CB(r)$). This gauge fixing condition is a consistent constraint with canonical quantization since the conjugate momentum $\pi$ to the field $\beta$ vanishes:
\begin{align}
\label{eqn:PrimaryConstraint}
    \pi=\frac{\partial{\mathcal{L}}}{\partial(\partial_t \beta)}=0.
\end{align}
Thus, \eqref{eqn:PrimaryConstraint} is a primary constraint that, together with the constraint $\beta=0$, form a pair of second-class type within Dirac's classification \cite{htbook}. We proceed with quantization by removing this pair of fields $(\beta,\pi)$ from the set of non-trivial operators that act on the underlying Hilbert space. We discuss this in details in \appref{app:DBQuant}. 
After removing that pair from the phase space, the conjugate momenta $P$ and $\Pi$ to $X$ and $\Theta$ are
\begin{align}
\label{eqn:XMomentum}
    P&=-\partial_r\Theta, \\
\label{eqn:ThetaMomentum}
    \Pi&=\partial_t\Theta,
\end{align}
and the Hamiltonian density
\begin{align}
\label{eqn:HamiltonianDensity}
    \CH=\frac{1}{2}\Pi^2+\frac{1}{2}\left(\partial_r X\right)^2,
\end{align}
where 
\begin{align}
\label{eqn:Hamiltonian}
    H=\int dr\, \CH.
\end{align}
Notice that the momentum conjugate to $X$ is given by $\partial_r \Theta$. This is reminiscent of the situation for open strings charged with respect to a background gauge field \cite{Abouelsaood:1986gd}, in that case the non-zero components of strings along different directions are canonically conjugate to each other. This connection is not that surprising once one recalls that the quantization, in that case, is based on combining components of strings into complexified fields. We comment more about this similarity in \cite{Albrychiewicz:2025hzt}. There we discuss the relationship between tropological sigma models and brane quantization, for which the quantization of the symplectic manifold is carried out by the quantization of the strings ending on branes that preserve A-model supersymmetry.

\section{Boundary Conformal Algebra}
\label{sec:Algebra}
We now consider a worldsheet with a boundary i.e., the strip given by \eqref{eqn:BdryWorldsheet}. We start with the discussion of the conformal symmetry generators at boundaries and their boundary algebra. Since we want to keep the boundary fixed at $r=0,\pi$, this requires that the generator $\delta r$ of \eqref{eqn:ConfSymGen1} vanishes there. With $r$ now being compact, we can decompose the infinitesimal generators $f(r)$ and $F(r)$ (in the process of analytic continuation we also redefined $F(r)=iF(r)$ in \eqref{eqn:ConfSymGen2} to preserve reality) into Fourier modes
\begin{align}
\label{eqn:BCFTgen}
    f(r)=\sum\limits_{m=1}^\infty\tilde{L}_m\sin(mr), \quad F(r)=\sum\limits_{m \in \Z}J_me^{imr},
\end{align}
where $f(r)$ vanishes at $r=0,\pi$ to preserve boundaries and the tilde on $L_m$ is used to distinguish these modes from the boundary-less case where
\begin{align}
\label{eqn:CFTgen}
    f(r)=\sum\limits_{m\in \Z} L_me^{imr}.
\end{align}

Without the boundary, the generators $L_m$ and $J_m$ satisfy the $\mathfrak{bms}_{2+1}$ algebra \cite{sachso,bondi,sachs,sachsgw,bicak} with two central charges $c$ and $c_\times$. This algebra was also shown to arise in the context of Carrollian field theories \cite{b,bi,bii,biii} and flat space holography \cite{Donnay:2022aba, Donnay:2022wvx}, 
\begin{align}
    [L_m, L_n]&=(m-n)L_{m+n}+\frac{c}{12}m(m^2-1)\delta_{m+n,0},\\
    [J_m, J_n]&=0,\\
    [L_m, J_n]&=(m-n)J_{m+n}+\frac{c_\times}{12}m(m^2-1)\delta_{m+n,0}.
\end{align}
Additionally, this algebra can also be derived as a Wigner-\.In\"on\"u contraction \cite{inonu1952} of two copies of the Virasoro algebra. 

We use the $L_m$ generators to construct the boundary version $\tilde{L}_m$. Comparing \eqref{eqn:BCFTgen} with \eqref{eqn:CFTgen}, we find that $\tilde{L}_m=-\frac{i}{2}(L_m-L_{-m})$. Using this relation, we derive a new boundary algebra with only one central charge surviving
\begin{align}
    [\tilde{L}_m,\tilde{L}_n]&=\frac{i}{2}\left((m-n)\tilde{L}_{m+n}-(m+n)\tilde{L}_{m-n}\right),\\
    [J_m, J_n]&=0,\\
    [\tilde{L}_m, J_n]&=-\frac{i}{2}\left((m-n)J_{m+n}-(m+n)J_{m-n}\right)+\frac{c_J}{12}m(m^2-1)(\delta_{m+n,0}+\delta_{-m+n,0}). 
\end{align}

\section{Boundary Conditions}
For a worldsheet with boundaries, the variation of fields for the action \eqref{eqn:TropStringsAction} gives total derivatives with respect to $r$ 
\begin{align}
    \int_{\partial\Sigma}dt\left\{-\delta X\partial_r X-\delta\Theta\partial_t X \right\},
\end{align}
that we can no longer ignore. Instead, to preserve the equations of motion, we impose boundary conditions. 

We discuss two options. One is a novel boundary condition that is a combination of Neumann and Dirichlet boundary conditions, nonetheless, we still refer to them as Neumann boundary conditions in this paper. They are imposed by
\begin{align}
\label{eqn:NeumannCond}
    \partial_r X|_{r=0,\pi}=0, \quad \partial_t X|_{r=0,\pi}=0. 
\end{align}
The second is closer to standard Dirichlet boundary conditions, where we fix the variation of fields at the boundaries
\begin{align}
\label{eqn:DirichletCond}
    \delta X|_{r=0,\pi}=0, \quad \delta \Theta|_{r=0,\pi} =0.
\end{align}
Since this has to be satisfied for arbitrary values of $t$, this implies
\begin{align}
    \partial_t X|_{r=0,\pi}=0, \quad \partial_t \Theta|_{r=0,\pi}=0.
\end{align}

One could also generalize these conditions by introducing background fields and pulling them back. For example, the introduction of a target space 1-form gauge field would modify the N boundary condition by adding terms of the form
\begin{align}
\label{eqn:WilsonLine}
    \int_{\partial\Sigma}dt\{A_X\partial_t X+A_\Theta\partial_t\Theta\},
\end{align}
to the variation of the action. We remark that the transition functions that define the gauge field have to be appropriately defined so that the action is invariant under foliation preserving diffeomorphisms and is consistent with the tropical geometry in the target space. 

\section{Open Strings Solutions and Canonical Quantization}
\label{sec:OpenStrings}

We construct open strings solutions of \eqref{eqn:Eom1}-\eqref{eqn:Eom3} using the boundary conditions derived above. With two distinct boundary conditions, one can also consider mixed boundary conditions for different fields. Here, we focus on the case where both directions have the same type of boundary conditions. First, consider the case where both boundaries have Dirichlet conditions for both $X$ and $\Theta$: we fix values of $X$ and $\Theta$ to be $x_0, x_1$ and $\theta_0,\theta_1$ at boundaries $r=0,\pi$ respectively. Then, the solutions satisfying \eqref{eqn:DirichletCond} can be written in the following form
\begin{align}
    \label{eqn:XModeExp}
        X(r,t)&=x_0+\frac{\Delta x}{\pi}r+\sum\limits_{n=1}^\infty X_n \sin(2nr), \\
    \label{eqn:YModeExp}    
        \Theta(r,t)&=\theta_0+\frac{\Delta \theta}{\pi}r+\sum\limits_{n=1}^\infty\Theta_n\sin(2nr)+t\sum\limits_{n=1}^\infty2n X_n(1-\cos(2nr)),
\end{align}
where $X_n$ and $\Theta_n$ are excitation modes and we defined $\Delta x=x_1-x_0$, $\Delta \theta=\theta_1-\theta_0$. Notice that in comparison to the relativistic open strings solutions, $X$ has no dependence on $t$, which is due to the equation of motion \eqref{eqn:Eom3}.  $\Theta$ depends on $t$ only up to a linear term, whose behavior itself is determined by the $X_n$ oscillation modes up to a constant.

To canonically quantize this theory, we start with the classical Poisson brackets
\begin{align}
    \left\{X(r,t),P(r',t)\right\}=\delta(r-r'), \quad \left\{\Theta(r,t),\Pi(r',t)\right\}=\delta(r-r'). 
\end{align}
Upon quantization, they get promoted to the following canonical commutation relations
\begin{align}
    \left[X(r,t),P(r',t)\right]=i\delta(r-r'), \quad \left[\Theta(r,t),\Pi(r',t)\right]=i\delta(r-r').
\end{align}
Using the solutions \eqref{eqn:XModeExp}, \eqref{eqn:YModeExp} together with the explicit form of conjugate momenta \eqref{eqn:XMomentum}, \eqref{eqn:ThetaMomentum} we find that the modes $X_n,\Theta_n$ satisfy the following commutation relations
\begin{align}
\label{eqn:ModeCan}
    \left[X_n,X_m\right]=0, \quad \left[X_n,\Theta_m\right]=\frac{i}{\pi}\delta_{n,m}.
\end{align}

From \eqref{eqn:HamiltonianDensity}, one observes that the Hamiltonian only depends on the modes $X_n$ that commute with each other and are already Hermitian. Using \eqref{eqn:Hamiltonian} we can find the Hamiltonian and it takes a simple form in terms of the Fourier modes,
\begin{align}
\label{eqn:HamiltonianDir}
    H=2\pi\sum\limits_{n=1}^{\infty}n^2X_n^2+\frac{\pi}{2}\left(\frac{\Delta x}{\pi}\right)^2.
\end{align}
The fact that $X_n$'s commute also implies that there is no need to regularize the ground states oscillation energies as in the relativistic case. In contrast to standard quantum field theory, which yields an infinite number of decoupled oscillatory modes, the theory presented here appears to produce a theory of infinitely many free particles. This outcome aligns with the original physical motivation behind the analytically continued tropological sigma models. These models describe the asymptotic behavior of the moduli space of punctured Riemann surfaces, where it is well known that strings become on-shell and propagate through a tower of increasingly massive physical states \cite{trsm}.

The form of the Dirac brackets \eqref{eqn:ModeCan} suggests that the modes $\Theta_n$ act as a conjugate momentum of the modes $X_n$. Therefore, we start by introducing states of the operators $X_n$ and construct the Hilbert space as a tensor product over all modes, for an arbitrary state we can write 
\begin{align}
    |X\rangle=|X_1\rangle\otimes|X_2\rangle\otimes|X_3\rangle\dots,
\end{align}
where each satisfies
\begin{align}
    X_n|X_n\rangle=x_n|X_n\rangle.
\end{align}
The vacuum is defined as a state for which $x_n=0, \; \forall n$. Since the Hamiltonian \eqref{eqn:HamiltonianDir} is diagonal in the operators $X_n$, the eigenstates $|X_n\rangle$ are also eigenstates of this Hamiltonian. Now, denoting eigenstates of $\Theta_n$ as $|\Theta_n\rangle$, where their construction is analogous to the one above, we can find the overlap to be
\begin{align}
    \langle X_n|\Theta_n\rangle= e^{-i\pi x_n\theta_n}.
\end{align}

Before, we consider solutions with different boundary conditions, we discuss the case where the string along $X$ is wrapped on a circle $S^1$ of radius $R$ by identifying points
\begin{align}
    \Delta x\rightarrow \Delta x + 2\pi m R,
\end{align}
with an integer $m$ being a winding number. This periodicity condition shifts the zero point of the Hamiltonian by 
\begin{align}
\label{eqn:HamWrapDD}
    \left(\frac{\Delta x}{\pi}+2mR\right)^2.
\end{align}

Consider another case where the fields $(X,\Theta)$ on both boundaries satisfy \eqref{eqn:NeumannCond}. Without any additional constraints, such boundary conditions only fix $X$. This is a significant difference in contrast to the relativistic case where the boundary conditions restrict all components along which they are enforced. For $\Theta$, only the $t$ dependent part of the solution is impacted by boundary conditions due to the equation of motion \eqref{eqn:Eom1}, which relates this part to $\partial_r X$ up to a constant $C$,
\begin{align}
    \label{eqn:XModeExpNN}
        X(r,t)&=x_0+\sum\limits\limits_{n=1}^\infty X_n \cos(2nr), \\
    \label{eqn:YModeExpNN}    
        \Theta(r,t)&=\theta_0+\sum_{n\neq 0}\Theta_ne^{i2nr}+t\left(C+\sum\limits\limits_{n=1}^\infty2n X_n\sin(2nr)\right).
\end{align}
As before, the zero mode of $X$ has no conjugate momentum since $X$ is time independent. 

The constant $C$ in the solution for $\Theta$ can be removed once we insert Wilson line type of terms into the boundary action integral and modify boundary conditions according to $\eqref{eqn:WilsonLine}$. The addition of boundary terms also changes the conjugate momentum and the quantization. If we keep $A_\Theta$ nonzero, the momentum conjugate to $\Theta$ \eqref{eqn:ThetaMomentum} is now given by
\begin{align}
    \Pi=\partial_t\Theta+A_{\pi}\delta(r-\pi)-A_{0}\delta(r),
\end{align}
where $A_{\pi}, A_0$ is the value of $A_\Theta$ on the boundaries $r=\pi,0$ respectively. The difference is denoted as $\Delta A=A_{\pi}-A_0$. Now the Hamiltonian is
\begin{align}
\label{eqn:HamWrapNN}
    H=2\pi\sum\limits_{n=1}^{\infty}n^2X_n^2+\frac{\pi}{2}\Delta A^2.
\end{align}
If we wrap $X$ on $S^1$ in this case, the Hamiltonian does not change since it does not depend on the momentum conjugate to the zero mode.

One could ask if there exists a tropical analog of T-duality. We would expect that the theory with $X$ field wrapped on $S^1$ and (DD) boundary conditions is dual to the theory with $X$ wrapped on $S^1$ with Wilson loops inserted and (NN) boundary conditions. However, the fact that $X$ in the latter case is already periodic implies that the Hamiltonian $\eqref{eqn:HamWrapNN}$ does not have ea term that is inversely proportional to $R$. We conclude that T-duality in its standard form is absent in this theory. This aligns with the intuition that tropical objects in a tropical geometry base correspond to geometric objects found in the A- and B-models. Therefore, one would not expect to find examples of mirror symmetry that descend to the underlying tropical geometry, as mirror symmetry is fundamentally a relation between the A- and B-models \cite{gross2011tropical}. This suggests that an appropriate tropicalization of the path integrals associated to the B-model would lead to the same tropological sigma models provided by a tropicalization of the A-model. We leave this as an open question for a future work.

\section{Conclusions}

In this paper, we reviewed the basics of a tropological A-model and discussed its analytic continuation. We examined the resulting theory on the worldsheet with boundaries and introduced a new type of boundary conformal algebra. Additionally, we derived Dirichlet and Neumann modified boundary conditions that arise in this theory and can be imposed at the endpoints of strings, leading to the tropical analog of a D-brane. We analyzed open string solutions with (NN) and (DD) boundary conditions and constructed their mode expansions.

Using canonical quantization, we derived commutator relations for the modes and provided a one-dimensional Hamiltonian, which is simpler than in the relativistic case due to the absence of regularization issues. The Hamiltonian depends only on commuting operators, and there is no energy shift caused by ground state oscillations. This Hamiltonian was interpreted as arising from the infinite tower of increasingly massive string states that one would expect to find when a string goes on-shell. With this Hamiltonian and the quantized operators, one can construct states and compute observables either directly or through an analog of conformal mapping, as in the standard relativistic case. We leave this exploration for a future work.

In comparison to the previous work \cite{trsm}, we demonstrated how tropicalization alters the quantization procedure for open strings. While the system is easier to solve compared to the relativistic case, several intricacies remain to be addressed. The differing boundary conditions for the fields $ X$ and $\Theta$ leave open the question of mixed boundary conditions. Additionally, we did not investigate the potential new dynamics that might emerge when the target space has a real dimension greater than two. As noted in \cite{trsm}, tropological sigma models allow generalizations to higher dimensions. Surprisingly, due to the nilpotency of higher-order Jordan structures $J$, these models naturally generalize to odd real dimensions where contact geometry becomes relevant. From $J$, one can construct actions and impose the boundary conditions discussed here to explore more exotic open string solutions.

Another interesting direction to explore is to discuss tropical D-branes that preserve A-model supersymmetry. In this case, one has to study a theory before the analytic continuation on the worldsheet with boundaries. There are in fact two possible types of branes that one can consider: Lagrangian and coisotropic branes \cite{Kapustin:2001ij} and their construction is dependent on the target space geometry. In \cite{Gukov:2008ve}, these objects are used for the procedure called brane quantization. This method approaches the quantization of a symplectic manifold $M$ by associating an A-model to the complexification of $M$. We discuss how this procedure changes in the tropical limit in \cite{Albrychiewicz:2025hzt}.

\acknowledgments
We would like to acknowledge that the program of tropological sigma models was originally initiated by Petr Ho\v{r}ava and we are thankful for his guidance. Additionally, we wish to thank Ori Ganor for the comments on the manuscript and Christopher J. Mogni and Jesus Sanchez Jr. for illuminating discussions and support. This work has been supported by the Berkeley Center for Theoretical Physics.

\appendix 

\section{Dirac-Bergmann's Quantization of the Tropical String Lagrangian}
\label{app:DBQuant}
Below we present the Dirac-Bergmann's quantization \cite{Bergmann:1949zz,Dirac:1950pj,Anderson:1951ta} of the tropical strings Lagrangian \eqref{eqn:TropStringsAction}. Recall that the Lagrangian is
\begin{align}
    \mathcal{L}=\frac{1}{2}\left(\partial_t \Theta\right)^2-\frac{1}{2}\left(\partial_r X\right)^2+(\beta-\partial_r\Theta)\partial_t X.
\end{align}
This system has two primary constraints
\begin{align}
\label{eqn:constraints}
    \phi_1&=\pi\approx 0, \\ \nonumber
    \phi_2&=P-(\beta-\partial_r\Theta)\approx 0,
\end{align}
where the conjugate momenta were defined as in \eqref{eqn:PrimaryConstraint}-\eqref{eqn:ThetaMomentum}. The total Hamiltonian is
\begin{align}
    H_T=\int dr \; \left(\frac{1}{2}\Pi^2+\frac{1}{2}(\partial_r X)^2+\lambda_1\phi_1+\lambda_2\phi_2\right),
\end{align}
with Lagrange multipliers $\lambda_1, \lambda_2$ to impose constraints \eqref{eqn:constraints} and $r$ is bounded in the interval $[0,\pi]$. Imposing canonical commutation relations at equal time (in this notation, we keep the time dependence of the fields implicit)
\begin{align}
    [X(r),P(s)]=\delta(r-s), \quad [\Theta(r), \Pi(s)]=\delta(r-s), \quad [\beta(r), \pi(s)]=\delta(r-s),
\end{align}
we find that constraints $\phi_1,\phi_2$ are of second class type since
\begin{align}
    [\phi_1(r), \phi_2(s)]=-\delta(r-s).
\end{align}
Next, we study the time evolution of constraints. Since we consider a theory with boundaries, we need to specify their type. In what follows, we use tropical Dirichlet boundary conditions \eqref{eqn:DirichletCond}, the analysis for different boundary conditions is analogous. 

With Dirichlet boundary conditions, we find
\begin{align}
    \dot{\phi}_1(r)=[\phi_1(r), H_T]=\int \;ds\lambda_2(s)\delta(r-s)\approx 0, 
\end{align}
which does not lead to another constraint but fixes the value of $\lambda_2=0$. Similarly, we compute
\begin{align}
    \dot{\phi}_2(r)=[\phi_2, H_T]=-\partial_r^2X(r)-\partial_r\Pi(r)-\lambda_1(r),
\end{align}
which fixes the value of the $\lambda_1$ multiplier. Therefore, there are no secondary constraints and we only need to address two constraints that are of second class type. 

The modified Dirac brackets are
\begin{align}
    [\Phi_i,\Phi_j]_D=[\Phi_i,\Phi_j]-[\Phi_i,\phi_k]C^{kl}[\phi_l,\Phi_j],
\end{align}
where $C^{kl}$ is an inverse of constraint matrix and $\Phi=(X,\Theta)$. We observe that
\begin{align}
    [\beta,\pi]_D=0,
\end{align}
confirming the claim from \secref{sec:AnalyticCont} that the conjugate pair $(\beta,\pi)$ should be removed from physical degrees of freedom. The Dirac brackets for conjugate pairs $(X,P)$ and $(\Theta, \Pi)$ are unchanged, and we may continue with quantization by choosing the appropriate, irreducible representation of operators using the mode expansion as described in \secref{sec:OpenStrings}.

\bibliographystyle{JHEP}
\bibliography{ticfpiqm.bib}

\end{document}